\documentclass[a4paper,floatfix,rmp,twocolumn,showkeys,superscriptaddress]{revtex4}
\usepackage{graphicx}
\usepackage{epstopdf}
\begin{document}

\title{Emergence of proto-organisms from bistable stochastic differentiation and adhesion}

\providecommand{\ICREA}{ICREA-Complex Systems  Lab, Universitat Pompeu
  Fabra,   Dr    Aiguader   88,   08003   Barcelona,   Spain}
\providecommand{\SFI}{Santa Fe  Institute, 1399 Hyde  Park Road, Santa
  Fe NM 87501, USA}
\providecommand{\IBE}{Institut de Biologia Evolutiva, UPF-CSIC, Psg Barceloneta 37, 08003 Barcelona, Spain}

\author{Salva Duran-Nebreda\footnote{joint first author}}     
\affiliation{\ICREA}   
\affiliation{\IBE} 

\author{Adriano Bonforti\footnotemark[\value{footnote}]}     
\affiliation{\ICREA}   
\affiliation{\IBE} 

\author{Ra\'ul Monta\~nez\footnotemark[\value{footnote}]}     
\affiliation{\ICREA}   
\affiliation{\IBE} 

\author{Sergi Valverde}     
\affiliation{\ICREA}   
\affiliation{\IBE} 

\author{Ricard Sol\'e\footnote{corresponding   author}}   \affiliation{\ICREA}
\affiliation{\SFI}
\affiliation{\IBE}
 
\vspace{0.4 cm}
\begin{abstract}
\vspace{0.2 cm}
The rise of multicellularity in the early evolution of life represents a major challenge for 
evolutionary biology. Guidance for finding answers has emerged from disparate 
fields, from phylogenetics to modelling and synthetic biology, but little is known about the potential 
origins of multicellular aggregates before genetic programs 
took full control of developmental processes. Such aggregates should involve spatial 
organisation of differentiated cells and the modification of flows and concentrations of 
metabolites within well defined boundaries. Here we show that, in an environment 
where limited nutrients and toxic metabolites are introduced, a population of cells 
capable of stochastic differentiation and differential adhesion can develop into multicellular aggregates 
with a complex internal structure. The morphospace of possible patterns is shown to be very 
rich, including proto-organisms that display a high degree of organisational complexity, far beyond simple 
heterogeneous populations of cells. Our findings reveal that there is a potentially enormous 
richness of organismal complexity between simple mixed cooperators and embodied living 
organisms.
\end{abstract}

\keywords{Multicellularity, Major Transitions, Complexity, Evolution}

\maketitle


\section{Introduction}


Multicellularity has evolved multiple times through the history of our planet, leading to a wide array of spatially 
organised living structures such as aggregates, sheets, clusters or filaments (1). 
The transition to multicellularity required the emergence of alternative cellular states along with stable, 
physical interactions among previously isolated cells (2-5). 
In our present-day biosphere, multicellular systems display intricate spatial and 
temporal patterns implemented by developmental programs, which are tightly controlled by genetic networks (6,7). But an early stage might have involved non-inherited stochastic phenotypic switches and physical aggregation phenomena that  
could have given rise to some class of cooperating multicellular assemblies (8,9). 
This is supported by the well-known observation that single-celled organisms can behave as multicellular systems
using precisely these processes (10) particularly in the face of high-stress events 
(11-13). Simple multicellular systems, such as {\em Anabaena}, 
where cell differentiation is induced under nitrogen deprivation, or mixobacteria (10)
are examples 
of the minimal types of multicellular organisation (14,15). A minimal form of 
multicellularity is provided by persister cells and phase variation 
phenomena, i.e. slow-growing cell subpopulations that can spontaneously 
switch back and forth among multiple resistant phenotypes, as a bet-hedging 
strategy in front of potential catastrophe(16,17). 

In this paper we aim to explore the potential for {\em organismality} 
(18) emerging from a minimal set of assumptions, 
including (a) multistability (19), incorporated as a stochastic 
bistable phenotype (20), allowing for two cell types ($1$ and $2$), 
(b) differential adhesion, which can lead to spatial segregation 
of different proto-tissues and pervades several key processes of development 
(5,21) and (c) a selective environment where the presence of external nutrient and toxic waste forces the selection of genotypes with higher fitness. 
Both types of cells can survive only in presence of nutrient, which is transformed into internal energy, and die if exposed to high concentrations of waste. Cells of type 2 have the additional capability of degrading waste in medium, at the expense of their capability of elaborating nutrient.
Previous models involving the evolution of 
undifferentiated multicellularity (22,23) have shown that 
appropriate metabolic trade-offs might pervade the coexistence of cell clusters. 
Our model goes a step further by allowing alternative cell states to organize in space. 
We find that if the system is allowed to exploit spatial organization, its evolution gives rise not only to cell heterogeneity, but also to nested substructures and to the creation of an internal environment, 
thus suggesting that combining differential adhesion and multistability provides the necessary toolkit for 
evolving proto-organisms in a robust manner.  
 The results reported here indicate that the generative potential which is typical of the morphological landscape can also be obtained by a simple, previously unexplored set of pattern-forming rules, where cell-cell communication or genetic networks are not taken in account.
 It contains the three key components of evolved MC (24) namely (a) fitness-coupled spatial patterning, (b) 
 cooperation and specialization and (c) a transition from "simple" to "complex" multicellular forms.

\section{Model Specifications}
Our assumption is that aggregative organisms involving multiple cell states present a better fitness than single-state 
organisms in a habitat with limited resources and toxic molecules. To delve into the accuracy of our assumption, 
we consider a model in which cells  are able to stochastically switch between two different metabolic states, and present differential 
cell-cell adhesion (Fig. 1). If a cell is able to survive to the habitat's conditions, it will spread its offspring, 
allowing to achieve maximum fitness by means of an evolutionary process involving mutation of parameters.

\subsection{Metabolism and competition}

A selective environment is introduced including both an incoming external nutrient ($N$) and a toxic waste
 ($W$) as well as an internal currency molecule ($E$). A regular $L \times L$ square lattice $(\Omega)$ is used. 
 Each site $(i,j) \in \Omega$ is characterized by a state, indicated as $S_{ij}$. This state 
 can be $0$ if the site is empty and either $1$ or $2$ if the site is occupied by cells. 
 These two values indicate two different cell types with different adhesion and metabolic properties. 
 Both $N$ and $W$ are added continuously to the empty lattice sites and 
 passively diffuse through the external medium and across nearest cells. 
 Energy $E$ is created by cellular metabolism, as an intracellular product of nutrient processing. 
 Cells of type 2 can allocate resources for waste degradation, at the cost of reduced nutrient elaboration, following a linear tradeoff 
 ($\varepsilon >0$) consistent with a maximum metabolic load and shared resources for protein synthesis. For type-1 cells we have $\epsilon=0$. 
 All three molecules experience linear degradation. 

The spatial dynamics are described by a discrete set of coupled 
differential equations. For each site $(i,j) \in \Omega$:
$${\partial N_{ij} \over \partial t} = \alpha I_{ij}
- (\eta_{N} + \pi_{ij}) N_{ij} + D_{N} \bigtriangledown^{2} N_{ij} $$ 
$${\partial E_{ij} \over \partial t} = \pi_{ij} N_{ij} - \eta_{E} E_{ij}$$
$${\partial W_{ij} \over \partial t} = \beta I_{ij} 
- \eta_{ij}  W_{ij}+ D_{W} \bigtriangledown^{2} W_{ij} $$
Here $\pi_{ij} = \rho \delta_{(1,S_{ij})} + \varepsilon \rho \delta_{(2,S_{ij})}$  and 
$\eta_{ij}=\eta_{W} + (1 - \varepsilon) \gamma \delta_{(2,S_{ij})}$  include decay and active removal of 
$N$ and $W$, respectively. We have used the Dirac's delta function $\delta_{kl}=1$ if $k=l$ and zero otherwise.  Similarly, the input terms $I_{ij}=\delta_{(0,S_{ij})}/\lambda$ for each site 
are effective provided that the site is empty. The normalisation factor $\lambda$ (the fraction of sites occupied by cells) 
ensures a constant flux of $N$ and $W$ throughout the lattice. 
A cell divides when its $E$-value increases beyond a fixed threshold ($\Theta_{div}$) and there is an empty site in the vicinity. 
This new cell inherits the genotype and the phenotypic state of the progenitor with a small chance of mutations (see SI), 
and the energy is equally split among the two (i.e. no asymmetric divisions are considered). 
Conversely, cells die if the $W$ value surpasses another fixed threshold ($\Theta_{tox}$), if $E$ falls 
below a critical value ($\Theta_{starv}$), or with a small random probability ($\xi$), releasing their contents 
($N$, $W$ and $E$ as nutrient) to the surrounding medium. Following this formulation, there is a natural 
competition for resources that can promote selection of different multicellular communities.  

\begin{figure}
\begin{center}
	\includegraphics[width=8.5 cm]{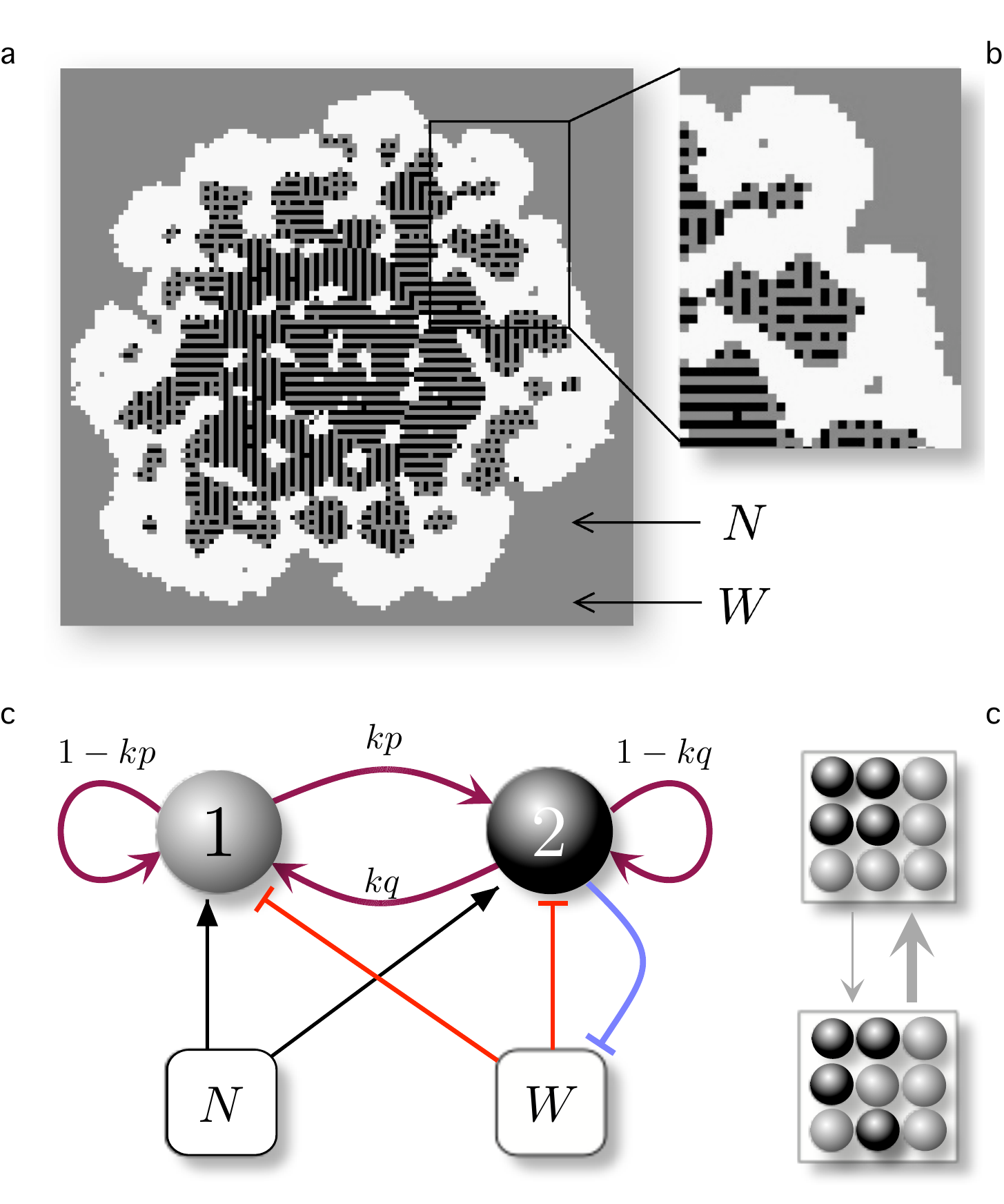}
	\caption{Minimal description of our model for the emergence of proto-organisms. The world is defined as a regular 2D lattice continuously seeded with nutrients and toxic metabolites in the empty space (white), where two types of cells (grey and black) can coexist. In (a) we show the example of an evolved proto-organisms displaying complex nested structures (b). (c) Cells can stochastically switch between the two available phenotypes, which can have different adhesive and metabolic properties. Phenotypic switching in both directions may occur with a certain probability (kp, kq), while ($1-kp$, $1-kq$) are the probabilities that no change occurs for grey or black cells respectively. Cells also interact with the local fields of a nurturing substance ($N$) and a toxic metabolite ($W$), which are involved in cell duplication, survival and death. The metabolism of black cell includes the ability to degrade waste (blue arrow). (d) In our model cells move by swapping locations with a neighboring cell, provided that the final energy is reduced, in accordance with their (evolved) adhesion properties.}
\label{Fig:fig_1}
\end{center}
\end{figure}

\begin{figure*}
\begin{center}
	\includegraphics[width=17 cm]{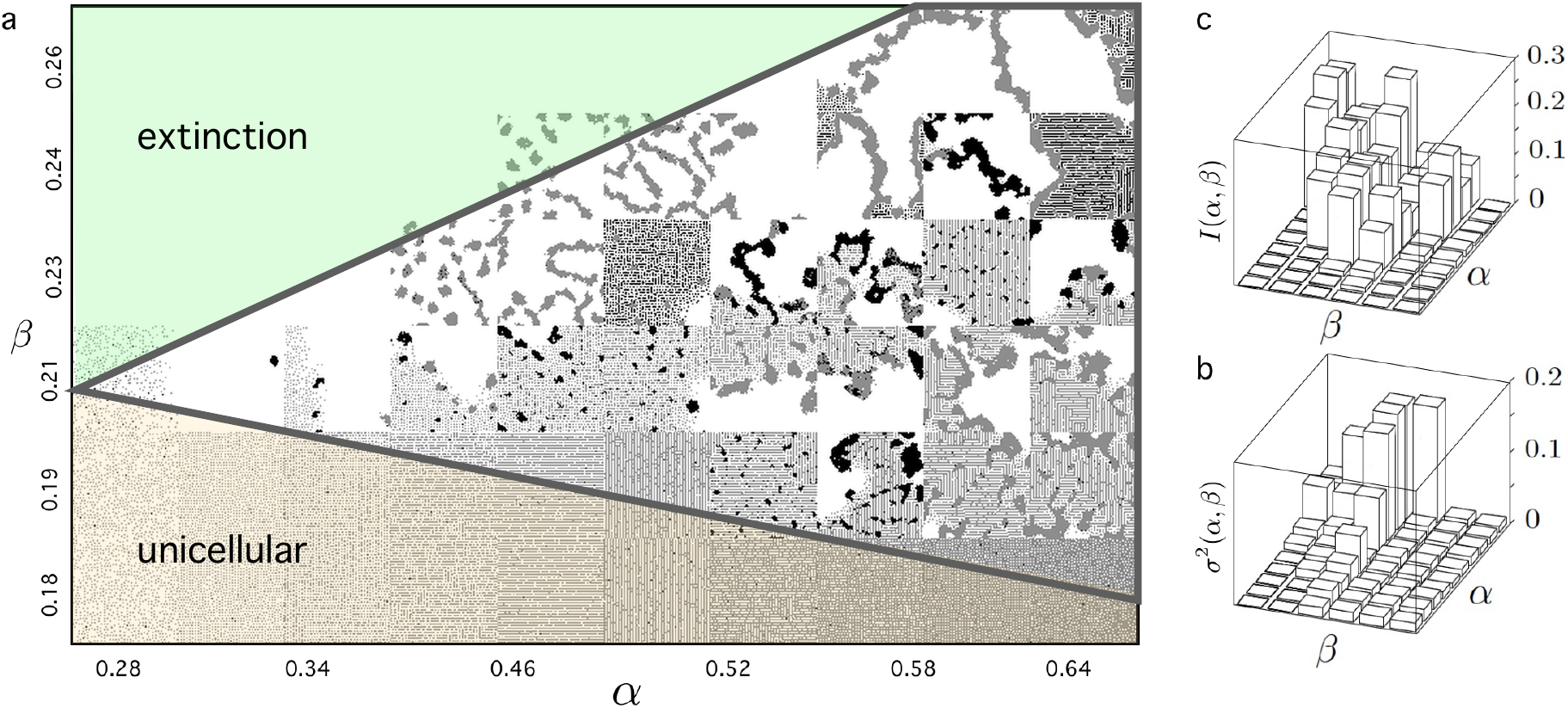}
	\caption{Impact of energy availability and waste input on the selected phenotypes. Here we show a representative 
	region of the system 	for different nutrient ($\alpha$) and waste ($\beta$) input values after $5\times10^5$ iterations of 
	the genetic algorithm (a). The multicellular region of this phase space is confined by a minimum energy density below which 
	aggregates cannot grow, but is otherwise expanded by increasing fluxes of both $N$ and $W$. Mutual information (b) of the 
	cellular states for neighbourhoods of size $3$ (see SI). Genetic diversity (c), calculated from the final normalized 
	genotypes (see SI). As $\alpha$ and $\beta$ become larger, more spatial structure is observed and increasingly 
	diverse genotypes are established, eventually settling to more than one species coexisting (see below).}
\label{Fig:fig_2}
\end{center}
\end{figure*}

\subsection{Stochastic switching genetics}

We introduce genetics in our model in the form of a stochastic transition between phenotypes (Fig. 1c) relevant for cell sorting and metabolism, 
as it is assumed in phase and antigenic variation in certain microbial populations (16,25,26). 
Specifically, cells can switch between states with evolvable probabilities $p, q \in [ 0,1 ]$, i. e.: 
$$P(1 \rightarrow 2) = \kappa p_{ij} \quad \quad \quad P(2 \rightarrow 1) = \kappa q_{ij}$$
where $\kappa$ is a fixed scaling factor, introduced to account for the time scale separation between adhesion kinetics and genetic processes. 
Therefore, the phenotypic transitions are not dependent on any molecular cue nor cellular memory beyond their current state. 

\subsection{Minimal model for cell adhesion}

The physics of cell sorting can be introduced considering the arrangement of cells 
constrained by their local preferences (27,28). Following Steinberg's differential adhesion model (DAH) 
we assume that cells movement are driven by the minimization of adhesion energy being cells more or less prone to remain together, avoiding the external medium, or maximizing contact with it (5,28-31).
An adhesion (or interaction) matrix ${\bf  {\cal J}}$ weights the strength of pairwise interactions among neigbouring sites:
$$
{\bf  {\cal J}} = 
\left( \begin{array}{ccc}
{\cal J}_{(0,0)} & {\cal J}_{(0,1)} & {\cal J}_{(0,2)} \\
{\cal J}_{(1,0)} & {\cal J}_{(1,1)} & {\cal J}_{(1,2)} \\
{\cal J}_{(2,0)} & {\cal J}_{(2,1)} & {\cal J}_{(2,2)} \end{array} \right)
$$ 
which is symmetric, i. e. ${\cal J}_{(a,b)}={\cal J}_{(b,a)}$, and has ${\cal J}_{(0,0)}=0$. Other approaches 
(32) consider each cell as formed by a number of sites, thus allowing for a better 
matching with the underlyng physics of cells. For simplicity, we keep our model confined to a one cell-one site 
scheme. 
Since cell-cell (and cell-medium) interactions are necessarily local (Fig. 1c), a given cell can only interact with a set $\Gamma_{ij}$
of eight nearest neighbors. The model allows cell movement between neighboring positions by switching the two local states 
provided that the final state is more likely to happen, i. e. consistent with 
the optimization of both cell adhesion energies. This is given by an energy function:
$$
{\cal H}_{ij} = \sum\limits_{kl \in \Gamma_{ij}} {\big[ {\cal J}_{(S_{ij}, S_{kl})}^{ij} + {\cal J}_{(S_{kl}, S_{ij})}^{kl}\big] \over (2 - \delta_{(0,S_{ij})} -\delta_{(0,S_{kl})})}
$$
which averages the interaction matrix of both cells. The superindexes denote the adhesion matrix of a particular site 
and normalise the effect of interacting with empty medium.

At each step, we choose a random neighbor for each site, compute the new 
energy ${\cal H}'$ and compare it to the original one ${\cal H}$. If the difference 
$\Delta {\cal H} = {\cal H}' - {\cal H}$ is negative, a decrease in the global energy would occur 
and thus the state swap is always applied. Instead, when $\Delta {\cal H}>0$, the largest the difference 
the less likely the change is assumed to happen, with a probability following the 
Boltzmann rule (for more details see SM1):
$$
 P(S_{ij} \rightarrow S_{kl}) = {1 \over 1 + e^{\Delta {\cal H}/T}}
 $$
As defined, the transition is likely to occur if an energy reduction takes place, with a noise factor 
introduced by $T$, acting as an effective "temperature". A small stochasticity prevents the system 
from getting trapped into local energy minima.

\section{Results}

\subsection{Resource and waste levels influence selection for complex multicellularity}

In order to analyse the prevalence of multicellular traits, we have explored the role 
played by nutrient and waste inputs in selecting different phenotypes by evolving 
the different parameters. The results are shown in Figure 2a. Simulations are started with type 1 cells only with no 
adhesion (i. e. , ${\cal J}_{(i,j)}=0$ for all adhesion strengths) thus behaving as random walkers (since $\Delta {\cal H}=0$ and thus 
$ P(S_{ij} \rightarrow S_{kl})=1/2$). This parameter space displays three main phases, including 
a cell-free (extinction) phase, a second phase with sparse distribution of unicellular populations (lower domain) and an intermediate 
phase (marked by a thick line) associated to organismal structures. 
Moreover, different measures were applied to these endpoint states of evolutionary processes (Fig. 2b-c), showing 
an overall increase in complexity for the multicellular region of this phase space in terms of structural organization and 
genetic diversity (see SI). In particular, the increase in genetic diversity is due to the existence of multiple distinguishable species
that create different, complex spatial arrangements, as measured by the spatial mutual information measure.

\subsection{Cellular embodiment enables niche construction}

Within the multicellular region of this phase space, proto-organisms display consistent 
spatial and temporal structures of remarkable complexity. Typically, an outer layer of cells that develops 
aggregative features in order to withstand the mounting 
levels of toxic waste, and which surrounds and protects an internal environment 
with lower $W$ and $N$ levels, suitable to be colonized by cells that preferentially expose to the environment. 
Within these "container" other cell types (not viable outside these boundaries) can coexist (Fig 1a). 
These nested structures define a proto-organism, reminiscent of biologically relevant organisations like $Volvox$ $sp.$, and 
can regenerate the protective layer in case it breaks or even create a whole new proto-organism, thus acting as a propagule (see SM1) 
and effectively defining a rudimentary life cycle.
Moreover, given the spatial constraints to the local concentration and flows of metabolites caused by the 
organisation of cell types, ecosystem engineering is also present (33,34). 

\begin{figure*}
\begin{center}
	\includegraphics[width=18 cm]{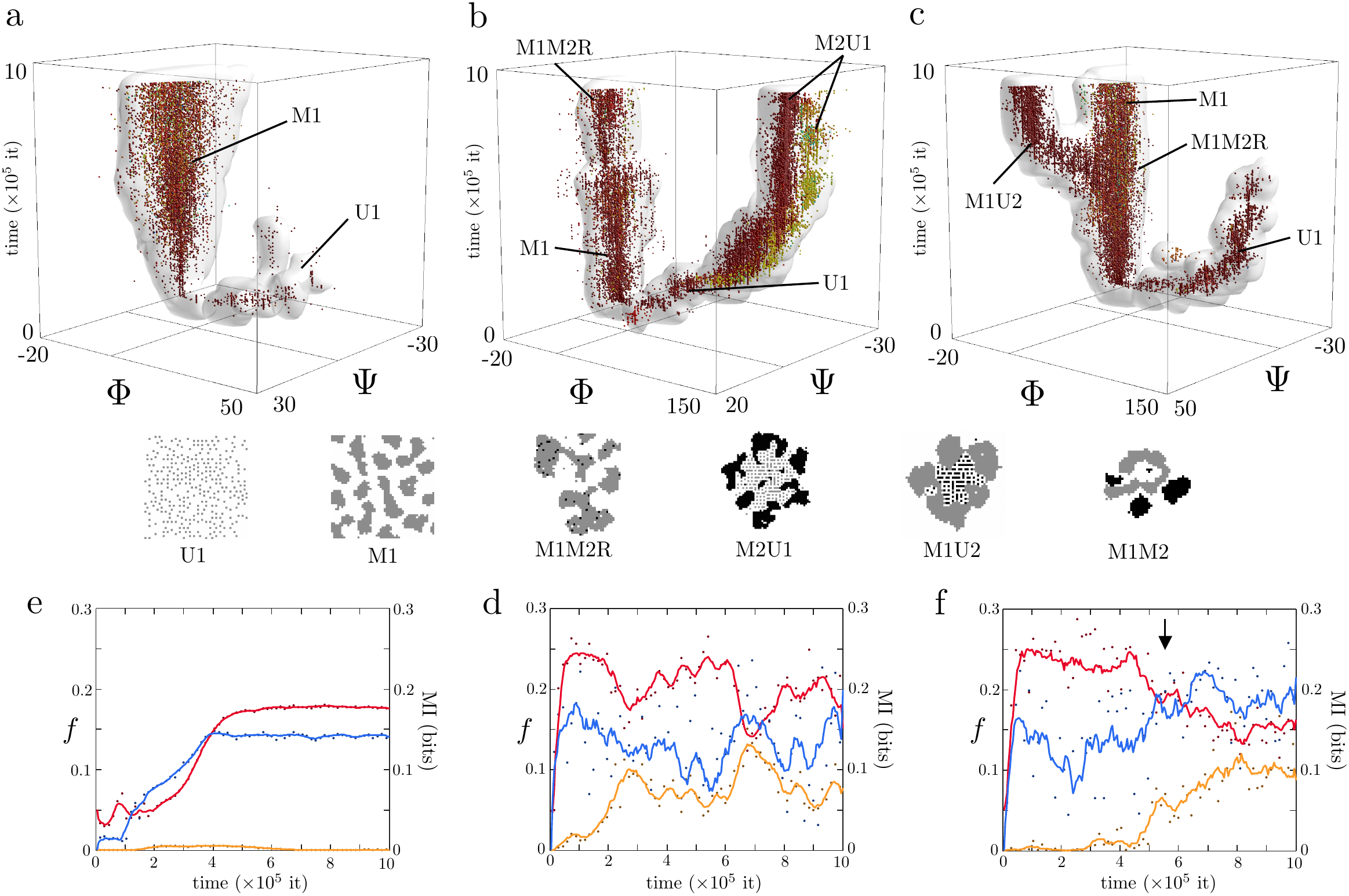}
	\caption{Speciation process in our GA. The plots on the first row 
	show the evolutionary dynamics of populations of cells in different parameter sets ($\alpha, \beta$), 
	forming distinct branches or ``species". Here, individuals are represented as dots in time coordinates ($z$ axis) as well as a reduced 
	genotype space ($x$, $y$, and colour). More precisely, we use the tendency of 
	grey cells to adhere to other grey cells $\Phi$ ($2{\cal J}_{(1,0)}-{\cal J}_{(1,1)} -{\cal J}_{(0,0)}$), 
	the tendency of black cells to adhere to other black cells $\Psi$ ($2{\cal J}_{(2,0)}-{\cal J}_{(2,2)}-{\cal J}_{(0,0)}$) 
	and the ability to process waste $\varepsilon$.
	An isosurface of genotype density is also shown (see SI). The lines on the bottom face of 
	each cube mark axes origin and the the starting population 
	genotype at $\Phi=\Psi=0$. As a guide for the eye, we provide a 
	representation of each of the main lineages at the bottom of the figure.
	On the second row, for each run we show the population dynamics of each cell type 
	in fraction of lattice occupied $f$ (red and yellow for types 1 and 2 respectively) 
	as well as the mutual information evolution in bits, right axis (blue). Points represent 
	actual values while lines are moving averages (binning window: 9 data points).}
\label{Fig:fig_3}
\end{center}
\end{figure*}

\subsection{Convergent evolution towards multicelullarity}

Beyond the small scale dynamics of the system, the particular paths taken by each population in the evolutionary process were also analyzed.
In figure 3 we display the evolutionary dynamics of three different scenarios using a reduced genotype space. 
In particular, we find that the tendency to form homo-aggregates of each cell type and the waste degradation potential 
yield a functional clustering of individuals into discrete subpopulations or species (see SI for a principal component analysis of the population genotypes). 
These populations will generate aggregates of a particular type if $2{\cal J}_{(a,0)}-{\cal J}_{(a,a)}-{\cal J}_{(0,0)} < 0$, and will display 
unicellular traits -i.e. will tend to attach to the external medium- otherwise. Also, cells will process more waste at the expense of efficiency in nutrient absorption 
the lower the $\varepsilon$ values.
 
The first example shows the evolution of a ``simple", undifferentiating aggregative species (M1) under medium energetic conditions 
and high inputs of waste. The other two -different runs of the same parameter set- display coexisting species, giving rise to complex 
multicellular phenotypes with differentiation and division of labor. Interestingly, all three cases share 
the same lineages for a short period at the beginning of the simulation yet soon diverge into different 
evolutionary histories. For instance, in the second scenario the type 1 unicellular lineage (U1) acquires a protective aggregative layer 
that is also proficient in processing waste (thus becoming M2U1), while in the third case it is the aggregative species M1 that fills the 
U1 niche once this strain disappears, evolving into M1U2 (see also SM2 3 \& 4). The evolution of these mirror multicellular proto-organisms M2U1 
(Fig. 3b) and M1U2 (Fig.c) -which are essentially the same phenotype with switched adhesion properties between the two 
cell types, yet coming from different lineages-, is a clear example of convergent evolution and path-dependance in our system. 
Figures 3e-f show the population dynamics and the evolution of the mutual information of the system in each simulation.
In the first case, the population follows a classic logistic growth and has a fixed, stable MI. The other two, instead, display heavy fluctuations in both the population levels and the mutual information, even showing signs of
quasi-periodic dynamics (b). It can also be clearly observed the emergence of a new species and its impact in the mutual information
in the third dataset (arrow marks the branching of the main species M1 into M1U2 at approximatelly $6\times10^5$ iterations).

\subsection{Collective fitness and epistatic interactions}

In previous examples some species are shown to coexist while others appear to be mutually exclusive, 
implying a rich repertoire of underlying ecological interactions. In order to better understand the fitness dependencies 
and evolution of multicellular traits in our model, we performed controlled experiments with some of the most commonly 
observed genotypes and two different environments: one with abundance of nutrient and waste, and a more stringent one with lower levels of nutrient and toxic metabolites.
In each simulation the lattice was inoculated with a few cells of one or two genotypes: 
namely $U1$, $M1$, $M1U2/M2U1$, $M1$ and $U1$ together, M1M2 and M1M2R.
Figure 4a shows the normalised population size in each scenario attained after $10^3$ iterations. Some genotypes appear to be viable in only
one of the two environments, while some (especially M1U2/M2U1) provide efficient growth in both, possibly
being the most fit genotype in fluctuating environments. 

 \begin{figure}[htp]
\begin{center}
	\includegraphics[width=8.5 cm]{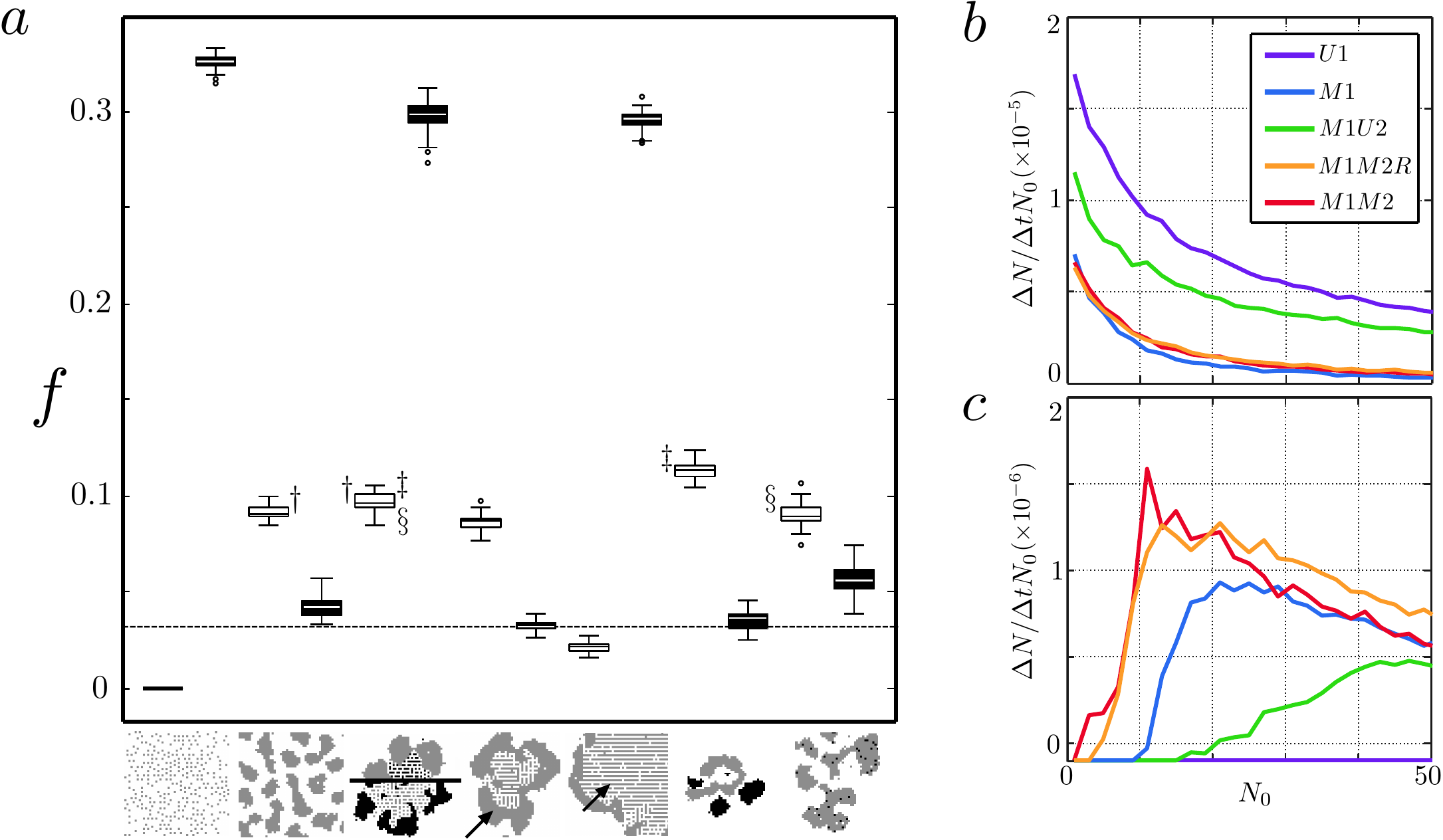}
	\caption{Fitness characterization of the most commonly observed species in the GA. 
	In (a) we show the occupied fraction of the lattice ($f$) after $10^3$ algorithm iterations of growth without mutation
	 by different genotypes or combinations of them. For two environments ($\alpha=0.36, \beta=0.193$ in filled boxes and 
	 $\alpha=0.56, \beta=0.247$ in white boxes), an empty lattice was inoculated with a few cells (basal dotted line) of one 
	 or two genotypes, from left to right: U1, M1, M1U2/M2U1, M1 and U1 together (M1 shown), M1 and U1 together (U1 shown), M1M2 and M1M2R. Boxes represent the limits of the first and the third quartile of the data, the central line is the median value and 
	 bars stand for the most extreme values not considered outliers, which are plotted individually as circles (50 simulations 
	 for each condition). $\S, \dagger,$ and $\ddag$ link datasets that were found to have different mean values (t-test, 
	 significance $p<10^{-4}$). In (b) and (c) we show the impact of the initial cell number ($N_0$) on growth rates 
	 ($\Delta N / \Delta t N_0$) for the same environments and genotypes previously used (25 simulations of each condition). 
	 In the high $W$ environment (c) a cooperative domain is found below a finite population size, in which adding cells to 
	 the propagule non-linearly increases growth rates. Beyond this fitness peak, cells compete for resources and space, 
	 leading to a cumulative decrease in fitness similar to the results obtained in the low $W$ environment (b).}	

\label{Fig:fig_4}
\end{center}
\end{figure}

Interestingly, the nature of the ecological interaction among genotypes U1 and M1 is shown to switch from straight competition in the low $W$ and $N$ scenario to parasitism in the high $W$ and $N$ scenario. In fact, in the first scenario both genotypes survive when alone but they compete for scarce resources when together, marking a decrease in the fitness of both genotypes. On the contrary, in the latter scenario only M1 survives when genotypes are alone, but creates a protective layer for U1 when genotypes live together, so that U1 can survive at the expenses of M1, which receives less $N$ and sees its fitness decrease. As commented previously, this is mainly caused by the 
capacity of multicellular entities to create an internal environment, which can be colonized by unicellular species, similarly to a parasitic 
microbiome-host relation. Figures 4bc, on the other hand, characterize the 
interactions between cells of the same genotype by varying the initial population 
size ($N_0$) in the same environments. Using the initial population and the growth 
after $10^3$ iterations we approximate the specific growth rate following: 

$$\mu N = {\delta N \over \delta t} \approx {\Delta N \over \Delta t}, \quad \quad \mu (N_0) \approx {\Delta N \over \Delta t N_0}$$

We observe that regardless of the genotype, cells compete for resources and space in the low energy input scenario (b), meaning 
that each cell added to the initial population decreases the growth rate of the whole. A very different set of interactions appears to be in place 
in the high waste environment (c), giving rise to a cooperation domain in which increasing 
the propagule size increases the growth rate of the whole, resorting afterwards to competition 
between cells. This suggests that, under this simple rules, an optimum propagule size exists 
and a fitness beyond the individual has emerged.

\section{Discussion}

Emerging multicellularity can be described as cooperative groups 
of cells assembled from independent replicators (18). This transition 
might have involved different paths, from mixed aggregates to clonal 
organisms with simple developmental plans and life cycles. The existing literature 
usually deals with cooperators achieving 
some kind of selective advantage as a consequence of mutualism, including 
spatial clustering or structured communities, as it occurs in 
biofilms (35-37). But true organismality, 
with a diverse set of cellular phenotypes arranged in space as a functional structure, 
has not been previously described as emerging from evolved interactions among simple 
virtual cells embodied as darwinian entities. 

Here we have provided a minimal set of rules grounded in biological processes 
that shift the selective pressures towards aggregative behaviour and division of labor.
Morphological complexity (38) increased throughout our simulations and a fitness
transfer was shown to be in place with the evolution of $\varepsilon$. Interestingly, those cells with lower 
values of $\varepsilon$, also displayed stronger attachment between them (lower $\Psi$). This suggests 
that from a game theoretical perspective cells become ``intelligent" players and try to surround 
themselves with other players whose strategy is the most mutually beneficial. Although our cells 
live in an non-clonal environment (39), they can manipulate who do they stand next 
to, potentially shaping local genetic relatedness. This would ensure that the investment in $W$ reduction 
mostly benefits cells with a similar genotype, paving the way for the evolution of cooperative and altruistic 
behavior (40).
 
As a premise for this model, we have assumed the existence of death promoting 
agents in the environment, of which there are several naturally 
occurring candidates, like: oxygen (41,42), 
secreted antibiotics (43) or exoenzymes and toxins (44,45). 
We think that a particularly interesting scenario is the one given by niche construction (46), 
in which the efforts to exclude extant microorganisms
from the population by other ecological players might drive the evolution of multicellularity.
Such relation would entail a coupling between ecological and organismal complexity,
a link that has eluded previous efforts in artificial life research (see (47) and references therein). 

By allowing our virtual cells to evolve through mutation of parameters affecting 
interactions with the external fields as well as with other cells, 
we provide a clear framework to evolve complexity under selection. 
The result of this is a system that spontaneously evolves, under many parametric 
conditions, to a complex, spatially organized multicellular state. 
The embodied structures emerging from the interplay of cell sorting and stochastic 
phenotypic switching display interesting and relevant features, including spatial 
modification of concentrations and flows of resources as well as temporal 
dynamics resembling proto-life cycles. In these spatial 
communities, pattern formation is enforced by the optimization of nutrient uptake 
along with an efficient removal of waste. In doing so, our cell 
assemblies arrange themselves in fitness-coupled collectives, indicating that organismality might be 
an inevitable outcome while solving the conflict associated to simultaneously dealing 
with both requirements. 

Our analysis also suggests that in the context of the evolution of organismality 
proposed in (18) our proto-organisms
would fit in the high cooperation-reduced conflict category. This class 
of entities harbours species of disparate complexity, yet all showing a 
fitness-relevant division of labor and differentiation, which stand at the 
core of our model. Specifically, some artificial organisms have been 
shown to include an exclusively cooperative domain 
and specialisation in terms of a metabolic trade-off, producing a fitness transfer from one cell type to
the other. Differentiation into terminal lineages (i.e. generating a soma), 
although not contemplated in our current formulation, appears to be a 
basic requirement to further reduce conflict among cells and attain ``true multicellularity".

The model presented here can be improved by incorporating a more realistic physics 
allowing for movement of aggregates (48) as well as heterogeneous 
media (49) where resources and waste might be generated in a non-homogeneous fashion. 
Similarly, we have limited ourselves to a binary switch, therefore confining the functional 
cell diversity to two main classes. We also assume that cell types are always 
alive, excluding the possibility of having material scaffolds formed through 
the differentiation processes, as it occurs with many solitary and colony-forming microorganisms in 
shallow waters. No less relevant in this context is the potential of creating multicellular systems 
by means of artificial evolution experiments (50-52) or 
synthetic biology approaches (53-56). There is a great potential 
associated to the use of existing genetic components to engineer pattern-forming modules. Our 
proposed minimal system might help defining feasible paths to implement proto-organisms.


\vspace{0.25 cm}

\noindent
{\bf Aknowledgments}

\vspace{0.25 cm}


We thank the members of the Complex Systems Lab for useful discussions. 
This work has been supported by the European Research Council Advanced grant, 
by the Bot\'in Foundation by Banco Santander through its
Santander Universities Global Division, a MINECO fellowship and by the Santa Fe Institute.

\vspace{0.25 cm}
\noindent
{\bf References}

\begin{enumerate}
\item
Knoll A. H. 2011. The multiple origins of complex multicellularity. 
{\em Annu. Rev. Earth Planet. Sci.} 39: 217-39.

\item
Bonner, J. T. 2001. {\em First signals: the evolution of multicellular development}. 
Princeton University Press. Princeton.

\item
Nedelcu, A.M. and Ruiz-Trillo, I. (eds.) 2015. {\em Evolutionary Transitions to Multicellular Life: Principles and Mechanisms}. Springer-Verlag, London.

\item
Rokas A. 2008. The Origins of Multicellularity and the Early History of the Genetic Toolkit For Animal Development. 
{\em Annu. Rev. Genet.} 42: 235-251.

\item
Forgacs, G., and Newman, S. A. 2005. {\em Biological physics of the developing embryo}. Cambridge U. Press, Cambridge.

\item
Carroll SB. 2001.Chance and necessity: the evolution of morphological complexity and diversity. 
{\em Nature} 409: 1102-1109.

\item
Erwin DH, Davidson EH \newblock (2009) The evolution of hierarchical gene regulatory networks.
\newblock \emph{Nat. Rev. Genet.} 10:141--148.

\item
Newman, S. A. and Baht, R. 2008. Dynamical patterning modules: physico-genetic determinants of morphological development and evolution.  {\em Phys. Biol.} 5: 015008.

\item
Newman, S.A., Forgacs, G. and M\"uller, G. B.  2006. Before programs: the physical origination of multicellular forms. 
 {\em Int. J. Dev. Biol.} 50: 289-299.

\item
Shapiro JA, Dworkin M (Eds) 1997. {\em Bacteria as Multicellular Organisms}. 
Oxford University Press, Oxford.

\item
Balaban NQ, Merrin J, Chait R, Kowalik L and Leibler S. 2004. Bacterial persistence as a phenotypic switch.
 {\em Science} 305:1622-1625.

\item
Lewis K 2007. Persister cells, dormancy and infectious disease. {\em Nat. Rev. Microbiol.} 5: 48-55.

\item
Lewis K 2010. Persister cells.  {\em Annu. Rev. Microbiol}. 64: 357-372.

\item
Zhang, C., Laurent S, Sakr, S., L. Peng and S. Bedu. 2006. Heterocyst differentiation and pattern formation in cyanobacteria: a chorus of signals.  {\em Mol. Microbiol}. {\bf 59}, 367-375.

\item
N. S. Wingreen and S. A. Levin. 2006. Cooperation among Microorganisms.  {\em Proc. Natl. Acad. Sci USA}. 4, 1486-1488.

\item
Henderson, I. R., Owen, P., and Nataro, J. P. 1999. Molecular switches: the ON and OFF of bacterial phase variation. 
 {\em Mol. Microbiol.}, 33(5): 919-932.

\item
Veening, J. W., Smits, W. K., and Kuipers, O. P. 2008. Bistability, epigenetics, and bet-hedging in bacteria. 
 {\em Annu. Rev. Microbiol.}, 62: 193-210.

\item
Queller DC and Strassmann JE. 2009. Beyond society: the evolution of organismality. 
 {\em Phil Trans R Soc B} 364: 3143-3155.

\item
Laurent M and Kellershohn N. 1999. Multistability: a major means of differentiation and evolution in biological systems. 
 {\em Trends Biochem Sci} 24: 418-422. 

\item
Eldar A and Elowitz MB 2010. Functional roles for noise in genetic circuits.  {\em Nature} 467: 167-173.

\item
Gumbiner GM. 1996. Cell adhesion: the molecular basis of tissue architecture and morphogenesis.  {\em Cell} 84: 345-357.

\item
Pfeiffer T, Schuster S and Bonhoeffer S. 2001. Cooperation and Competition in the Evolution of ATP-Producing Pathways. 
 {\em Science} 292: 504-507.

\item
Pfeiffer T and Bonhoeffer S. 2001. An evolutionary scenario for the transition to undifferentiated multicellularity. 
 {\em Proc. Natl. Acad. Sci USA}. 100: 1095-1098.

\item
Niklas KJ and Newman SA. 2013. The origins of multicellular organisms. {\em Evol. Dev.} 15: 41-52.

\item
Hallet, B. 2001. Playing Dr Jekyll and Mr Hyde: combined mechanisms of phase variation in bacteria. 
\em{Current opinion in microbiology}. 4(5), 570-581.

\item
Darmon, E., and D.R.F. Leach. 2014. Bacterial Genome Instability.{\em Microbiology and Molecular Biology Reviews} : MMBR 78 (1): 1?39.

\item
Steinberg, M. S. 1964. The problem of adhesive selectivity in cellular interactions. 
In: {\em Cellular membranes in development} Vol. 22, pp. 321-366. Academic Press, New York.

\item
Foty, R. A., and Steinberg, M. S. 2005. The differential adhesion hypothesis: a direct evaluation. 
 {\em Dev. Biol.} 278(1): 255-263.

\item
Steinberg, M S. 1975. Adhesion-Guided Multicellular Assembly: A Commentary upon the Postulates, Real and Imagined, of the Differential Adhesion Hypothesis, with Special Attention to Computer Simulations of Cell Sorting. 
 {\em J. Theor. Biol.} 55 (2): 431-43.
 
\item
Hogeweg, P. 2000. Evolving Mechanisms of Morphogenesis: 
on the Interplay between Differential Adhesion and Cell Differentiation. 
{\em J. Theor. Biol.} 203: 317-333 

\item
Goel, N, R D Campbell, R Gordon, R Rosen, H Martinez, and M Ycas. 1970. Self-Sorting of Isotropic Cells. 
 {\em J. Theor. Biol.} 28 (3): 423-68. 

\item
Glazier, J. A. and Graner, F. 1993. Simulation of the differential adhesion driven rearrangement of biological cells. 
Emergence of multicellularity in a model of cell growth, death and aggregation.  {\em Phys. Rev. E} 47: 2128-2154.

\item
Jones, C. G. , Lawton, J. M. and Shachak, M. 1994. 
Organisms as ecosystem engineers. {\em OIKOS} 69: 373-370. 

\item
Erwin, D.H. 2008.  Macroevolution of ecosystem engineering, niche construction and diversity. 
{\em Trends Ecol Evol.} 23: 304-310.

\item
Branda, S. S., Vik, A., Friedman L and Kolter R. 2005. Biofilms: the matrix revisited. 
Trends Microbiol. 13: 20-26.

\item
Battin T.J, Sloan, W.T., Kjelleberg S. et al 2007. Microbial landscapes: new paths to biofilm research. 
{\em Nat. Rev. Microbiol.} 5: 76-81.

\item
Nadell C.D., Bucci V., Drescher, K. et al 2012. Cutting through the complexity of cell collectives. 
{\em J. Roy. Soc. Interface} 280: 20122770.

\item
Valentine JW, Collins AG, and Meyer CP. 1994.  Morphological complexity increase in metazoans. 
 {\em Paleobiology} 20: 131-42.
 
\item
Queller, D. C. 2000. Relatedness and the fraternal major transitions.  
{\em Phil. Trans. R. Soc. London B} 355: 1647-1655.

\item
West, S., I Pen, and AS Griffin. 2002. Cooperation and Competition between Relatives. {\em Science} 296 (5565): 72?75.

\item
Schirrmeister, B. E., de Vos, J. M., Antonelli, A., and Bagheri, H. C. 2013. Evolution of multicellularity coincided with increased diversification of cyanobacteria and the Great Oxidation Event. {\em Proc. Natl. Acad. Sci USA.} 110(5),

\item
Johnston, D. T., Poulton, S. W., Goldberg, T., Sergeev, V. N., Podkovyrov, V., Vorobeva, N. G., Bekker, A. and Knoll, a. H. 2012. Late Ediacaran redox stability and metazoan evolution. {\em Earth and Planetary Science Letters}. 335-336, 25-35.

\item
Ben-Jacob, E., Cohen, I., Golding, I., Gutnick, D. L., Tcherpakov, M., Helbing, D., and Ron, I. G. 2000. Bacterial cooperative organization under antibiotic stress. 
{\em Physica A: Statistical Mechanics and Its Applications}. 282(1-2), 247-282.

\item
Pattus, F., Massotte, D., Wilmsen, H. U., Lakey, J., Tsernoglou, D., Tucker, A., and Parker, M. W. 1990. Colicins: prokaryotic killer-pores. {\em Experientia}, 46(2), 180-192.

\item
Hibbing, ME, C Fuqua, MR Parsek, and SB Peterson. 2010. Bacterial competition: surviving and thriving in the microbial jungle. {\em Nat. Rev. Microbiol.} 8 (1): 15?25. 

\item
Odling-Smee, F. J., Laland, K. N., and Feldman, M. W. 2003. {\em Niche construction: the neglected process in evolution}. 
Princeton University Press. Princeton.

\item
Sol\'e R.V. and Valverde, S. 2013. Macroevolution in silico: scales, constraints and universals. 
{\em Paleontology} 56: 1327-1340.

\item
Sol\'e R.V. and Valverde, S. 2013. Before the endless forms: 
embodied model of transition from single cells to aggregates
to ecosystem engineering. {\em PLoS One}, 8: e59664.

\item
Rainey PB, Travisano M. 1998. Adaptive radiation in a heterogeneous environment. {\em Nature} 394: 69-72.

\item
Boraas, M., Seale, D., and Boxhorn, J. 1998. Phagotrophy by a flagellate selects for colonial prey : A possible origin of multicellularity. 
{\em Evolutionary Ecology} 1973: 153?164. 

\item
Ratcliff, W. C., Denison R. F., Borrello M. and Travisano M. 2012. Experimental evolution of multicellularity. 
{\em Proc. Natl. Acad. Sci USA.} 109: 1595-1600.

\item
Duran-Nebreda, S. and Sol\'e, R.V. 2015. Emergence of multicellularity in a model of cell growth, death and aggregation under size-dependent selection.  {\em J. Roy. Soc. Interface}. 12102.

\item
Basu S, Gerchman Y, Collins CH, Arnold FH, Weiss R. 2005. A 
synthetic multicellular system for programmed pattern formation. {\em Nature} 434: 1130?1134.

\item
Maharbiz M.M. 2012. Synthetic multicellularity. {\em Trends Cell Biol} 22: 617-623.

\item
Davies, J. A. 2008. Synthetic morphology: prospects for engineered,
self-constructing anatomies. {\em J. Anat.} 212: 707-719.

\item
Chuang, J. S. 2012. Engineering multicellular traits in synthetic microbial populations. 
{\em Curr. Opin. Chem. Biol.} 16: 370-378.
\end{enumerate}

\end{document}